# From Intracellular Signaling to Population Oscillations: Bridging Scales in Collective Behavior

Allyson E. Sgro[1,2], David J. Schwab[1,2], Javad Noorbakhsh[3], Troy Mestler[1], Pankaj Mehta[3], and Thomas Gregor[1,2]

[1]Joseph Henry Laboratories of Physics, Princeton University, Princeton, NJ 08544
[2]Lewis-Sigler Institute for Integrative Genomics, Princeton University, Princeton, NJ 08544
[3]Department of Physics, Boston University, Boston, MA 02215

**ABSTRACT**

Collective behavior in cellular populations is coordinated by biochemical signaling networks within individual cells. Connecting the dynamics of these intracellular networks to the population phenomena they control poses a considerable challenge because of network complexity and our limited knowledge of kinetic parameters. However, from physical systems we know that behavioral changes in the individual constituents of a collectively-behaving system occur in a limited number of well-defined classes, and these can be described using simple models. Here we apply such an approach to the emergence of collective oscillations in cellular populations of the social amoeba *Dictyostelium discoideum*. Through direct tests of our model with quantitative *in vivo* measurements of single-cell and population signaling dynamics, we show how a simple model can effectively describe a complex molecular signaling network and its effects at multiple size and temporal scales. The model predicts novel noise-driven single-cell and population-level signaling phenomena that we then experimentally observe. Our results suggest that like physical systems, collective behavior in biology may be universal and described using simple mathematical models.

# INTRODUCTION

Collective behavior is a common feature of many biological systems and is present in systems ranging from flocking birds, human spectators, schooling fish, and circadian rhythms in many higher organisms, to swarming bacterial colonies, cell migration, and embryonic morphogenesis (Ballerini et al, 2008; Couzin & Krause, 2003; Farkas et al, 2002; Friedl & Gilmour, 2009; Giardina, 2008; Kawano et al, 2006; Szabó et al, 2006; Ullner et al, 2009; Waters & Bassler, 2005; Zhang et al, 2010). In cellular systems that exhibit collective behavior, individual cells must coordinate their behavior with one another to produce the observed population-level phenomena and do so utilizing extracellular small molecules or proteins. For example, bacteria commonly utilize quorum-sensing molecules to synchronize gene expression in cellular populations and form aggregate biofilms (Waters & Bassler, 2005) and synthetic biology has exploited these mechanisms to engineer new circuits that give rise to population-level behaviors (Mondragón-Palomino et al, 2011; Youk & Lim, 2014). However, each cell's behavior and its communication with other cells is controlled by complex intracellular biochemical networks. Illuminating how the dynamics of these intracellular networks lead to the population-wide collective behavior observed in these systems is a challenging problem, in part due to the difference in size and temporal scales at which these behaviors are controlled and displayed.

A classic example of such collective cellular behaviors is the transition during starvation from an independent, single-celled state to a multicellular aggregate in the eukaryotic social amoeba *Dictyostelium discoideum*. Its population-level behaviors are controlled by a complex biochemical network within individual cells and coordinated through cell-cell communication via the small molecule cyclic AMP (cAMP). During starvation, single cells begin to produce cAMP and up-regulate the expression of signaling network components. Once this signaling machinery is sufficiently expressed, cells can detect the external cAMP and respond by massively producing their own pulse of cAMP internally that is then released into the environment. Released cAMP diffuses through the extracellular environment, relaying the stimulation to other cells and eventually leading to autonomous population-level oscillations (Alcantara & Monk, 1974; Gregor et al, 2010; Gross et al, 1976; Tomchik & Devreotes, 1981).

Despite our extensive knowledge of the components of the *Dictyostelium* signaling pathway, there is no consensus on how this pathway gives rise to synchronized cAMP oscillations in cellular populations (Laub & Loomis, 1998; Lauzeral et al, 1997; Martiel & Goldbeter, 1987; Sawai et al, 2005). Not only are there likely other as-yet-undiscovered components in the signaling circuit, the circuit dynamics are poorly understood and these dynamics change with increasing starvation time and changing environmental conditions. As a result, it is challenging to uncover the origins of collective behavior and predict novel behaviors even in a well-studied model organism such as *Dictyostelium* through a detailed, "bottom-up" modeling approach that incorporates each network component and interaction. These challenges are made even more pronounced by the need to bridge multiple time scales. For example, chemotactic responses to cAMP in *Dictyostelium* occur on the order of 30-60 seconds (Iglesias & Devreotes, 2008; Iglesias & Devreotes, 2012; Manahan et al, 2004; Takeda et al, 2012; Wang et al, 2012), whereas the period of population-level cAMP oscillations are typically an order of magnitude larger (6-10 minutes) (Gregor et al, 2010; Tomchik & Devreotes, 1981). Another approach is clearly needed to elucidate how these complicated single-cell networks give rise to collective population behaviors, and to bridge the divide between different temporal and size scales.

Here we present a general modeling approach for overcoming these challenges based on the concept that population-level behaviors do not depend on all details of the intracellular dynamics of individual members of the population and that a dimensionally reduced system accurately captures the essential phenomena. This approach has been key to understanding collective behavior in physical systems such as, for example, the equilibrium phase transition from a gas to a liquid (Anderson, 1997; Guckenheimer & Holmes, 1983; Kadanoff, 2000). In dynamical systems, at transitions between two different behaviors, so-called *bifurcations*, only a few qualitatively different behaviors are possible and these can be described by simple, low-dimensional models regardless of the complexity of the system being modeled (Izhikevich, 2007; Strogatz, 2001). This concept is known as "universality" (Kadanoff, 2000).

Many biological systems also undergo bifurcations in their behavior and may thus be amenable to mathematical modeling via a universality-based approach. For example, both isolated *Dictyostelium* cells as well as cellular populations undergo a bifurcation to oscillations as a function of external cAMP levels (Gregor et al, 2010; Tomchik & Devreotes, 1981). Here, we exploit universality to build a simple predictive model of the *Dictyostelium* signaling circuit that reproduces the essential behavior of single cells as well as cellular populations and experimentally confirm its success. This "top-down" modeling approach does not require detailed knowledge of the signaling circuit and is ideally suited for complex biological regulatory networks where kinetic or topological information is limited. Using this approach, we show that a universal model can successfully describe both single-cell and multicellular dynamics in collective biological systems, such as oscillatory cell populations of amoebae or neurons.

**RESULTS**

**A 2D-model for *Dictyostelium* signaling dynamics**

Population-level *Dictyostelium* signaling dynamics have been experimentally described in great detail (Gregor et al, 2010; Laub & Loomis, 1998; Martiel & Goldbeter, 1987; Sawai et al, 2005) but a comprehensive model that captures the basic phenomenology and yet retains predictive power is still missing. Thus, guided by experimental observations, our goal is to build a low-dimensional single-cell model, experimentally test its predictions, and then use it as a building block for a model that describes population-level signaling dynamics.

The key experimental observation underlying our single-cell model is a qualitative change, or bifurcation, in the *Dictyostelium* signaling network's dynamical behavior in response to increasing concentration of extracellular cAMP in a microfluidic device, measured using a FRET sensor (Figure 1 A-B, Figure E1, Figure E2) (Gregor et al, 2010; Nikolaev et al, 2004). At low extracellular cAMP levels, cells respond by producing a single pulse of internal cAMP whereas at high cAMP, cells oscillate. Hence, the extracellular cAMP concentration plays the role of a bifurcation parameter, and the system should be describable by a simple, low-dimensional model (Izhikevich, 2007; Strogatz, 2001). Furthermore, low-dimensional dynamical systems close to a bifurcation can exhibit only a few universal, qualitative behaviors, often termed bifurcation classes. The existence of small, noisy sub-threshold fluctuations in the baseline of internal cAMP levels even in the absence of extracellular cAMP, and the fact that varying extracellular cAMP is sufficient for the bifurcation to occur implies that the *Dictyostelium* signaling network is well described by a co-dimension one bifurcation (i.e. only one parameter needs to be varied for the bifurcation to occur), which is the simplest bifurcation class consistent with oscillations.

The simplest two-dimensional model that satisfies the above conditions is the excitable FitzHugh Nagumo (FHN) model (FitzHugh, 1961; Izhikevich, 2007; Murray, 2007; Nagumo et al, 1962). The FHN model falls into the supercritical Hopf bifurcation class, where a single stable fixed point transitions into a stable limit cycle that increases in amplitude as the bifurcation parameter increases (Figure E3). It has been studied extensively in the theory of dynamical systems and neuroscience (Izhikevich, 2007), and formally has oscillations that rise from zero amplitude and finite frequency with no bistability. Specific to the FHN model, when its bifurcation parameter is pulled rapidly upward while remaining below the bifurcation threshold, it exhibits a transient spike-like response with a characteristic amplitude, analogous to when neurons produce a voltage spike and fire. Because of that analogy, we adopt these terms from neuroscience to describe cells producing an internal pulse, or spike, of cAMP. Other models of *Dictyostelium* signaling dynamics based on excitability have been proposed, but most differ in fundamental ways. The FHN model is unique in its reliance on internal feedbacks, as we emphasize below.

The model has two dynamical variables: an "activator", $A$, which activates itself through an auto-regulatory positive feedback and a "repressor", $R$, that is activated by $A$ and, in turn, inhibits $A$ through a slower negative feedback loop (see Figure 1C). Mathematically, the noisy FHN is described by the stochastic Langevin equations

$$\frac{dA}{dt} = f(A) - R + I([cAMP]_{ex}) + \eta(t) \tag{Eq 1}$$

$$\frac{dR}{dt} = \varepsilon(A - \gamma R) + c_0 \tag{Eq 2}$$

where the non-linear function $f(A) = A - \frac{1}{3}A^3$ mimics the effect of a positive feedback loop. The constitutive activator degradation $\varepsilon$ controls the ratio between the activator and repressor time-scale dynamics, i.e. the excitability; $\gamma$ is the repressor degradation rate, and $c_0$ sets the steady-state repressor value in the absence of external cAMP. The input function $I([cAMP]_{ex})$ depends on the experimentally-controlled extracellular cAMP concentration, $[cAMP]_{ex}$, and reflects any 'pre-processing' modules that may exist upstream of the excitable FHN circuit. Experimentally, we find that the upstream 'pre-processing' circuit senses fold-changes in cAMP (Figure 1D) and thus is well modeled by $I(x) = a \log\left(1 + \frac{x}{K_d}\right)$. $K_d$ corresponds to the threshold for response to cAMP and $a$ determines the magnitude of the response (see SI of Sawai, et al. 2005). We have also included a Langevin noise term $\eta(t)$ that satisfies the relation $\langle \eta(t)\eta(t')\rangle = \sigma^2 \delta(t - t')$, where $\sigma^2$ is a measure of the strength of the noise. Importantly, due to universality our qualitative predictions do not depend strongly on the choice of parameters and the form of the non-linearity of $f(A)$ (Strogatz, 2001). All we require is that in the absence of external cAMP, the system is below the oscillatory bifurcation and excitable. Nonetheless, parameters were chosen to best fit the experimental data and a single set of parameters was used throughout this manuscript. Figures 1E and 1F show the phase portraits for a FHN model for externally applied cAMP stimuli (E) below and (F) above the threshold for oscillations, respectively. In the low cAMP phase portrait (Figure 1E), the final fixed point is stable and describes the long-time behavior of the system (Figure 1G). In contrast, in the high cAMP phase portrait (Figure 1F), the fixed point is unstable and the trajectories converge on a limit-cycle attractor for non-linear oscillations (Figure 1H). Thus, the activator concentration $A$ is a good proxy for the experimentally-observed intracellular cAMP levels, allowing for facile comparison between model and experiments.

One of the prominent behaviors of the FHN model is that in response to steps of external cAMP below the threshold for oscillations (Figure 1E), the trajectory makes a long excursion through phase-space resulting in a spike of the activator. This excursion produces a transient spike in the "internal cAMP" levels analogous to those seen in experiments (Figure 1A). Such spikes have also been observed previously where this behavior was interpreted as "adaptation" of the adenylyl cyclase, ACA, responsible for production of intracellular cAMP in response to changes in extracellular cAMP levels (Comer & Parent, 2006). In contrast, our model here indicates that these so-called "accommodation spikes" result directly from the underlying excitability of the intracellular *Dictyostelium* signaling circuit. Accommodation spikes occur frequently in models of dynamical systems and the terminology is borrowed from firing neurons, emphasizing here the connection between these vastly different systems.

**Single *Dictyostelium* cells are excitable feedback systems**

Before using this model as a building block for describing cellular populations, we performed a series of experimental tests concentrating on qualitative predictions of our dynamical model that do not depend on the detailed choice of parameters. Our model predictions for the time dependence of activator $A$ are well matched to our experimental data for single-cell cytosolic cAMP responses to externally applied cAMP stimuli (Figure 1A, 1B, 1G, and 1H). Notice that the model reproduces the initial accommodation spikes for all values of externally applied cAMP followed by oscillations for the 10 μM stimulus. We do find that some longer-term behaviors, such as a slight dampening of the oscillations or down-regulation of noisy firing, do not exactly match our phenomenological model as it is lacking additional terms that would represent such long-term behavior. This is due to genetic regulation becoming a factor in our experiments at these longer (> 10 min) timescales. However, the uniformly elevated levels of external cAMP applied here are unnatural and our model correctly reproduces single-cell responses to all naturalistic stimuli.

Over a wide range of cAMP concentrations in experiments, accommodation spikes quickly increase to their peak value, but differ in their decay time back to baseline (Figure 2A). Both in our model simulations and experimental data, accommodation spike widths monotonically increase with increasing extracellular cAMP concentration (Figure 2B and 2C), but the period of the ensuing oscillations for stimuli of 100 nM cAMP and above decreases as extracellular cAMP concentration increases (Figure 2D and 2E). Together, these results confirm that our model accurately represents the internal cAMP dynamics governing both the initial accommodation spike and the subsequent oscillatory behavior, suggesting that the molecular mechanism underlies both phenomena. This demonstrates that although single-cell oscillations are not observed in natural conditions, we can make predictions about natural behaviors from modeling their existence. Note that the observed scaling of accommodation spike widths scales logarithmically with increasing cAMP (Figure 2B), further validating our choice of a logarithmic preprocessing module. Furthermore, the spike width scaling is also inconsistent with the dynamics of adaptation by an incoherent feedforward network such as those proposed to govern *Dictyostelium* chemotaxis (Takeda et al, 2012; Wang et al, 2012). These results confirm that the full cAMP signaling circuit is best described using a feedback mechanism, and provide evidence that the observed dynamics result from an underlying negative-feedback architecture.

**Single cells are sensitive to the rate of stimulus change, not threshold sensors**

One of the most interesting predictions of our model is that internal cAMP responses depend on the rate of externally applied cAMP levels (Figure 3A and 3B) (Levine et al, 1996; Sawai et al, 2005). In particular, the model predicts that the signal propagation circuit will respond with an accommodation spike in response to a sub-oscillation threshold step of external cAMP (e.g. 1 nM), but will show no response for a slow ramp to the same external cAMP level (Figure 3A). The underlying reason for this difference in responses to a step vs. a ramp is best understood through the phase portrait (Figure 3A and 3C). For a sufficiently slow ramp of stimulus, the dynamics of the system can follow the stable fixed point as it moves through phase space, without ever leaving its equilibrated state, analogously to a thermodynamic system that follows a slow temperature change adiabatically. In contrast, for fast changes in external cAMP levels, the dynamics are no longer adiabatic and the large sudden change in position of the stable fixed point elicits an accommodation spike. Our experimental results are in agreement with our simulated model predictions (Figure 3A), showing that cells are sensitive to the rate of change of stimulus, and are in direct contrast to previous model assumptions that single cells behave as binary threshold sensors that spike as soon as a certain extracellular cAMP concentration is achieved (Levine et al, 1996; Sawai et al, 2005). Similar behavior was recently experimentally observed in the stress-response of *Bacillus subtilis* (Young et al, 2013).

A second test of our model is the response of the *Dictyostelium* signal propagation circuit to a large exponential ramp of externally applied cAMP that transitions the system from a sub- to a super-oscillation threshold level (e.g. 100 pM to 300 nM). Our model predicts that once the external cAMP levels are increased adiabatically beyond a critical threshold value where the fixed point changes stability, the system will start oscillating with the amplitude of the oscillations growing with increasing externally applied cAMP (Figure 3B and 3D). Once again, our experimental results are in agreement with the simulated model prediction (Figure 3B). This behavior validates our initial assumptions that the *Dictyostelium* signaling circuit is in the supercritical Hopf bifurcation class and can be described by the FHN model (Strogatz, 2001).

**Single cell properties bound multicellular behaviors**

As a final test of our single-cell model, we probed how single cells respond to repetitive stimulation of the type seen during collective oscillations. While excitable systems respond to small changes in inputs, they also have a large refractory period where they become insensitive to further stimulation after the stimulus is withdrawn, not merely after its original onset. This interplay between the ability to respond to pulses, immediately followed by a refractory period, can be probed by subjecting cells to pulses with different widths and subsequent rest periods. Experimentally, single cells can easily be entrained to short (1 minute) pulses of external cAMP with a long (5 minute) rest period between each pulse. However, when the pulse width is increased to 5 minutes with only a minute of rest between each pulse, poor entrainment results (Figure 4A). We quantified single cell responses to a range of pulse widths and pulse periods, measuring entrainment quality as the mean correlation between the first period response and subsequent period responses (Figure 4B). As pulse widths approach the full length of the period, there is insufficient rest time for the system to relax to its previous steady-state and thus poor entrainment results. Similarly, our model simulations produce a well-entrained response to a variety of pulse widths, but only after a sufficient refractory period has passed (Figure 4C).

Entrainment is a natural experimental test of the *Dictyostelium* signaling network. During the aggregation stage of development, cells detect waves of external cAMP similar to the pulses in our entrainment experiments. Our experiments indicate that a pulse of 3-4 minutes must be followed by a 2-3 minute refractory period for entrainment to occur. These observations are consistent with estimates of the cAMP wave widths and periods found in aggregating populations (Tomchik & Devreotes, 1981). Furthermore, as shown in Figure 2A, the 3-4 minutes pulse width seen in aggregating populations can naturally arise from single-cell accommodation spikes in response to a wide range of external cAMP inputs. Together, these results suggest that the excitability of individual cells places strong limitations on the dynamics of collective oscillations in *Dictyostelium* populations.

**Population model reproduces critical behaviors**

Thus far we have theoretically reproduced a wide range of single-cell behaviors. However, to extend our mathematical description to cellular populations and their autonomous, synchronized oscillations, an explicit model for the dynamics of extracellular cAMP is necessary. We capture these dynamics by a simple mean-field approach, meaning that cells and stimulant are well mixed, thus neglecting spatial detail. Our multi-cell model builds on the basic structure of the adapted FHN model above where each cell $i$ is described by an activator $A_i$ and a repressor $R_i$ (Equations 1 and 2, respectively). Cells secrete low levels of cAMP at a constant baseline rate $\alpha_0$, and extracellular cAMP is degraded at a rate $D$. In addition, when a cell spikes, it releases a large cAMP pulse into the environment at a rate $S$. Finally, experimentally we can also flow additional cAMP into the system at a concentration $\alpha_f$. These behaviors are captured by a system of Langevin equations of the form

$$\frac{dA_i}{dt} = f(A_i) - R_i + I([cAMP]_{ex}) + \eta_i(t) \tag{Eq 3}$$

$$\frac{dR_i}{dt} = \varepsilon(A_i - \gamma R_i) + c_0 \tag{Eq 4}$$

$$\frac{d[cAMP]_{ex}}{dt} = \alpha_f + \rho\alpha_0 + \rho S \frac{1}{N}\sum_{i=1}^{N} \Theta(A_i) - D[cAMP]_{ex}, \tag{Eq 5}$$

where $\eta_i(t)$ is a cell-dependent Gaussian white noise term with $\langle n_i(t)n_j(t')\rangle = \sigma^2\delta_{ij}\delta(t-t')$, and $\rho$ is the cell density by volume. Equation 5 describes the spike-driven secretion of cAMP into the medium by the sum, $\rho S \frac{1}{N}\sum_{i=1}^{N}\Theta(A_i)$ with $\Theta(A_i)$ being the Heaviside function, which is equal to 1 if $A_i > 0$ and 0 otherwise. Note that the degradation rate $D = J + \alpha_{PDE}\rho$, where $\alpha_{PDE}$ is the basal rate of phosphodiesterase secretion, can be modulated in our experimental set-up by changing the flow rate $J$ in our microfluidic devices (see Materials and Methods).

Our model reproduces the phase diagram for a wide range of environmental conditions from Gregor et al. (2010), indicating that our mean-field approach is sufficient to describe the autonomous, synchronized population-level oscillations exhibited in *Dictyostelium* populations (Figures 5A and 5B). Note that these collective oscillations are actually synchronous accommodation spikes, not the induced oscillations resulting from an unstable limit cycle as with the single-cell oscillations. Plotting the population firing rate in terms of $\rho/J$ results in a data collapse for sufficiently large J, and we find good agreement between the model-predicted firing rate dependence on $\rho/J$ (Figure 5C) and the previously published data (replotted in Figure 5D). The ability of our model to accurately reproduce observed population-level behaviors over a wide range of experimental conditions demonstrates the success of our strategy of using the carefully calibrated single-cell model as a building block for a multi-cell description.

When the flow rate $J$ is sufficiently high the external cAMP concentration dynamics become fast, leading to a separation of timescales between individual cell dynamics and external cAMP dynamics. As a result, external cAMP can be thought of as quickly reaching a quasi-steady-state. This assumption dramatically simplifies our analysis of the model, because it allows us to ignore the dynamics of the external medium. Conceptually, it is helpful to think of the extracellular cAMP as originating through two distinct processes: the "firing-induced cAMP", $\frac{\rho S}{J}$, which measures the cAMP released by cells when they spike, and the "background cAMP", $\frac{\alpha_f + \rho\alpha_0}{J}$, which is the cAMP present even when cells do not spike. Together, these two sources of cAMP are sufficient to describe the external medium and constitute the basis variables for computing single-cell and population firing rates (Figure 6A). The single-cell firing rate is a measure of how often an individual cell in the population fires, averaged over the population; the population firing rate is a measure of how often the population fires synchronously. In a coherent population these measures produce similar results, whereas in an incoherent population the population firing rate vanishes. Together, these two quantities provide a succinct way to summarize the behavior of the system and allow us to determine in which areas of phase space cells are oscillating and whether these oscillations are synchronous or asynchronous.

To probe the predictive power of our population-level model, we performed a series of simulations where collectively oscillating cellular populations are subject to step-stimuli of additional external cAMP into the medium (Figure 6B). Considering the phase diagrams in Figure 6A, the addition of extracellular cAMP increases the parameter $\alpha_f$ and hence moves the system *horizontally* in the phase diagram, in contrast to changing the cell density or external flow rate which moves the system *diagonally* through the phase diagram. Without direct access to extracellular cAMP concentrations, *a priori* it is unclear where natural populations are located in this phase diagram. One possibility is that collectively oscillating *Dictyostelium* populations reside in "knee region" of Figure 6A labeled Point I. If so, flowing in a small amount of additional extracellular cAMP to these populations will lead to slowing of the synchronized population-level oscillations (Point II in the phase diagram) and a further increase in the amount of extracellular cAMP will result in a loss of oscillations both at the level of single cells and cellular populations (Point III). These behaviors are in stark contrast to that of single cells where flowing in additional cAMP always increases the firing rate. The model also suggests that adding even more external cAMP to the system (Point IV) will cause individual cells to start oscillating asynchronously. In this regime, individuals have a high firing rate but fire incoherently resulting in a negligible populating firing rate.

To test these predictions, we probed these autonomously oscillating populations of *Dictyostelium* with varying levels of additional cAMP at multiple flow rates (see Figure 6C and 6D). For extremely low levels of added extracellular cAMP, the cellular population continues to oscillate synchronously. However, when we increase the cAMP levels further (~2-10 nM), the oscillations slow down. Eventually, collective oscillations disappear for intermediate levels of cAMP (10 nM-20 nM). Finally, when the extracellular cAMP concentration is increased to extremely high levels ($\geq$100 nM) there is a marked increase in baseline level of the population-level FRET signal, indicating unsynchronized, autonomous oscillations of single cells. The ability to continue to synchronize oscillations over low background levels of cAMP (~2-10 nM) allows populations to maintain collective states in environments where degradation of secreted cAMP is never complete.

**Intracellular noise drives population-level phenomena**

The agreement between our simulated phase diagram and population-level experiments in Figures 5 and 6 suggests that our phenomenological model correctly captures the key biological mechanisms that give rise to collective oscillations in *Dictyostelium* populations. However, the question of what are the key biological mechanisms beyond a positive feedback loop remains unanswered. Previous work on autonomously oscillating *Dictyostelium* populations strongly suggests noise may also be key to the onset of collective oscillations (Gregor et al, 2010). To elucidate the role of noise in the intracellular cAMP circuit in the emergence of these collective oscillations, we recomputed our phase diagram for various levels and sources of noise in Equation 3 (see Figure E4). In the absence of stochasticity (Figure 7A and Figure E4B) or for noise solely in the external cAMP levels (Figure E4C), the phase diagram loses almost all of the structure seen in Figure 6A. Specifically, the knee region where synchronously oscillating *Dictyostelium* naturally reside without additional external cAMP (i.e. Point I) disappears in the single-cell phase diagram and the simulation results are inconsistent with the data collapse seen in Figure 5. However, the firing rate of individual cells near the oscillatory bifurcation depends strongly on noise, and population-level oscillations emerge in our model when one or a few cells stochastically spike and drive the rest of the population into synchrony. Therefore intracellular noise introduces a form of "stochastic heterogeneity" among identical cells that drives collective synchronization. To test whether similar effects could also arise from cell-to-cell heterogeneity in the levels of extracellular cAMP needed to induce a cytosolic spike, we repeated the simulations with a mixture of cells representing two cellular populations with different parameters. We found that the resulting phase diagram again has a different shape than for a homogenous, noisy population and is insufficient to reproduce the observed population behaviors (Figure E4D). These simulations indicate that in the presence of small amounts of extracellular cAMP heterogeneity alone cannot drive synchronized population oscillations.

In the simulations, for regimes with intermediate and high extracellular cAMP levels synchronized population spikes no longer occur. In these regimes our model predicts that stochasticity drives unsynchronized spiking, and that that spiking increases with increasing external cAMP levels (Figure 6A and 6B). Testing this prediction experimentally lets us verify directly whether intracellular noise indeed drives the observed population dynamics. To observe the presence of unsynchronized spiking, we measured cytosolic cAMP levels over time for single cells in autonomously oscillating populations that were subjected to external cAMP steps (Figure 7B). We find that in populations the standard deviation of single-cell cytosolic cAMP levels increases with increasing external cAMP levels (Figure 7C). This roughly 21% increase in the standard deviation over time in conjunction with the increased population-mean cAMP levels indicates that the single-cell levels of cytosolic cAMP themselves are noisier with increased spike height and/or frequency as external cAMP levels increase. While few clear accommodation spike-like cAMP peaks appear to be present after external cAMP is applied (Figure 7B), noise-driven spiking may not always result in full-size cAMP peaks like those that result from natural cAMP waves or step stimuli. For example, while a 1 minute 1 nM external stimulus is sufficient to produce a full-size peak of internal cAMP in a single cell such as those shown in Figures 1-4, external cAMP stimuli that are shorter than 20 second produce smaller peaks of internal cAMP (Figure 7D) that are qualitatively similar to those seen in Figure 7B. Together, our model and experiments suggest that *Dictyostelium* populations not only exploit stochasticity in their underlying intracellular signaling network to initiate collective population-level behaviors, but also to coordinate them in noisy extracellular environments.

**DISCUSSION**

We have developed a new conceptual framework for understanding collective behavior in cellular populations based on universality and used it for analyzing the emergence of collective oscillations in the social amoebae *Dictyostelium discoideum*. Our model accurately predicts and reproduces a plethora of experimentally-confirmed complex dynamical behaviors despite having minimal knowledge about the kinetic parameters and interactions of the underlying circuit. Our studies revealed that the dynamics of the *Dictyostelium* signaling network can be understood using a simple two-variable model, specifically the noisy and excitable FitzHugh-Nagumo model. We showed that individual cells are sensitive to the dynamics of the input signal and respond differently to steps and ramps of extracellular cAMP. We also showed that the excitability of individual cells leads to entrainment properties that fundamentally constrain the dynamics of population-level oscillations. When we extend our model to cellular populations, synchronized oscillations spontaneously arise from stochastic accommodation spikes, suggesting that *Dictyostelium* cells actively exploit stochasticity in the biochemical network for controlling population-level behaviors.

Our simple model explains a number of disparate biological phenomena observed during the initiation of the collective phase of the *Dictyostelium* life cycle. For example, it has been shown that when subjected to steps of external cAMP, adenyl cyclase A, which is responsible for cAMP synthesis, shows an initial peak of activation followed by a period in which its activity subsides even in the presence of stimulus (Comer & Parent, 2006). We have shown that this behavior naturally arises from the excitability of the *Dictyostelium* signaling circuit and is the analogue of what has been termed "accommodation spikes" in the neuroscience literature (Izhikevich, 2007). Furthermore, it was shown that mutants lacking phosphoinositide 3-kinases PI3K1 and PI3K2 no longer exhibit this accommodation spike behavior, and continually produce cAMP (Comer & Parent, 2006). The data from these experiments is consistent with the idea that the mutants undergo a bifurcation to oscillation at lower levels of extracellular cAMP. Our model also suggests a natural explanation for why *Dictyostelium* produce extracellular phosphodiesterases (PDEs) to degrade extracellular cAMP (reviewed in Saran et al., 2002). In the absence of degradation, the dynamics of the extracellular cAMP can no longer track the intracellular dynamics of cells, resulting in a loss of coherent oscillations. This loss of coherence is a generic phenomenon present in oscillator systems that communicate through an external medium (Schwab et al, 2012a; Schwab et al, 2012b). This is consistent with experiments on mutants showing that cells lacking extracellular PDEs do not give rise to spiral waves (Sawai et al, 2007) as well as the loss of population oscillations in the presence of the PDE inhibitor DTT (Gregor et al, 2010).

In addition to resolving these disparities in our understanding of *Dictyostelium* signaling dynamics, our model also allows us to discriminate between possible signaling network architectures. Currently there is no consensus on the architecture of the *Dictyostelium* signaling circuit downstream of the CAR1 receptors (Bagorda et al, 2009; Kimmel & Parent, 2003; Laub & Loomis, 1998; Sawai et al, 2005; Takeda et al, 2012; Wang et al, 2012). Our results indicate that there must be a negative feedback loop that turns off production of intracellular cAMP and that there is an as-yet-undiscovered intracellular positive feedback in the circuit. In contrast with recent work showing that upstream of cAMP production, the *Dictyostelium* chemotaxis network and its output is well-described by a feedforward network architecture (Takeda et al, 2012; Wang et al, 2012), our cAMP signaling circuit uses a network architecture based on feedback. Thus, the *Dictyostelium* chemotaxis and signal propagation networks differ in their underlying functional topologies (feedforward versus feedback), even though they likely have shared components and interactions. Indeed, this feedforward network may be the source of the logarithmic preprocessing we observe. One reason for this discrepancy is that the chemotaxis and signal propagation networks operate at fundamentally different time scales. Whereas the gradient sensing response is measured to be on the order of 30 seconds (Iglesias & Devreotes, 2008; Iglesias & Devreotes, 2012; Manahan et al, 2004), the signaling network response operates on a timescale that is nearly an order of magnitude larger (Gregor et al, 2010; Tomchik & Devreotes, 1981). For example, the chemotactic response of the pleckstrin homology (PH) domain of cytosolic regulator of adenylyl cyclase (CRAC) dynamics occur in tens of seconds and not in minutes as is the case of internal cAMP (Wang et al, 2012). This difference in timescales likely reflects different biological functions: the chemotaxis network is designed primarily to climb shallow gradients whereas the signal propagation network is designed to allow cells to aggregate into multicellular structures.

While FHN-inspired models have been previously used to describe population aggregation and cAMP waves in aggregates and slugs (Lee, 1997; Marée et al, 1999; Vasiev et al, 1997; Vasiev et al, 1994), the mechanism underlying collective oscillations in our coupled FHN population model differs fundamentally from earlier mathematical models for collective oscillations in *Dictyostelium*. Many classical models of *Dictyostelium* oscillations posit receptor desensitization as the primary mechanism through which oscillations emerge (Martiel & Goldbeter, 1987). Our model is similar in spirit to integrate-and-fire models in neuroscience, with cells only spiking once their internal state reaches a threshold, as opposed to the input reaching a threshold before spiking as was the case in previous models (Levine et al, 1996; Sawai et al, 2005). While these input-threshold models are successful at describing population-level behaviors, they are inconsistent with the dynamics we observe at the single-cell level (Figure 3A). For this reason, it is notable that the FHN model is able to correctly reproduce the observed single-cell dynamics and yet describe population-level dynamics as successfully as these input-threshold models.

A fundamental feature of our model that makes this accurate description of experimentally-observed behaviors possible is stochasticity. The work presented here suggests that *Dictyostelium* cells actively exploit stochasticity in the signal propagation network to initiate and control population-level oscillations, and while more examples of this type of exploitation are coming to light it is still clear that in many biological systems noise limits or degrades biological function (Eldar & Elowitz, 2010; Pilpel, 2011; Sanchez et al, 2013). Specifically, the increased likelihood of stochastic spiking at higher extracellular cAMP levels suggests a possible mechanism behind the origin of autonomous oscillation centers. The slow leakage of cAMP during development postulated by Gregor et al. (2010) will likely not induce accommodation spikes directly as cells are sensitive to the rate of extracellular cAMP change (Figure 3A), but the extracellular cAMP that builds up will encourage increased stochastic spiking. These initial stochastic spikes release comparatively large amounts of cAMP into the extracellular environment, triggering other cells to spike and driving the cells that fired first to become the origins of autonomous oscillation centers. This same stochasticity in the intracellular signaling network drives cells to fire when background levels of cAMP do not continually reset to zero, allowing for continued collective oscillations at times when complete degradation of extracellular cAMP does not occur.

Our work offers a bridge between the disparate fields of collective cellular signaling and neuroscience. The FHN model successfully describes the dynamics of both neurons and *Dictyostelium* cells. Neurons use electrical impulses on the order of a millisecond to communicate directly via synapses. In contrast, communication in *Dictyostelium* cells occurs indirectly through the external medium and is mediated via phosphorelays and second messenger molecules such as cAMP over several minutes. Despite these different architectures and timescales, due to the underlying universal dynamics both systems can be described by the same model and exhibit qualitatively similar behaviors including noise-induced accommodation spikes. This analogy suggests that many of the novel behaviors seen in neurons are likely to have an analogue in *Dictyostelium* populations, an observation that may be an interesting route for further exploration. Our work suggests that, like physical systems, collective behavior in biology may be universal and well described using simple mathematical models. Universality has played a fundamental role in furthering our understanding of physical systems, and we suspect it will also play an important role in furthering our knowledge of collective behavior in biology.

## MATERIALS AND METHODS

**Cell Culture, Preparation, and Genetic Manipulation**
Axenic *Dictyostelium discoideum* cell lines expressing Epac1camps (AX4 background, gift of Dr. Satoshi Sawai), Epac1camps and mRFPmars (AX4 background), ECFP, and EYFP (both AX3 background, gifts of Dr. Carole Parent) were grown according to standard protocols (Fey et al, 2007). Briefly, vegetative cells were grown at 22 °C while shaking at 180 rpm in PS medium consisting of 1.0% special peptone (Oxoid), 0.7% yeast extract (Oxoid), 1.5% D-glucose, 0.14% $KH_2PO_4$, 0.012% $Na_2HPO_4$-$7H_2O$, 40 ng/mL vitamin B12, 80 ng/mL folic acid, and 1X antibiotic-antimycotic mix (Gibco) supplemented with 5 µg/mL (EYFP/AX3), 10 µg/mL (Epac1camps/AX4), or 20 µg/mL (ECFP/AX3) G418. Vegetative cells were washed and shaken at 1-2x$10^7$ cells/mL in development buffer (10 mM K/$Na_2$ phosphate buffer, 2 mM $MgSO_4$, and 0.2 mM $CaCl_2$ for 4-5 hours prior to experiments.

The expression vector pBSRH-mars, which permits constitutive expression of mRFPmars in *Dictyostelium discoideum* under control of the *act15* promoter, was kindly provided by Dr. Robert Cooper. The Epac1camps strain was transformed with pBSRH-mars by electroporation following a standard protocol (Gaudet et al, 2007) and a clone was selected based on fluorescence intensity. Fluorescence intensity of mRFPmars appears uniform in the cytosol.

**Microfluidic Device Fabrication**
Microfluidic devices for both single cell and population experiments were fabricated using standard photolithography techniques to generate silicon masters and standard poly(dimethylsiloxane) (PDMS) replica molding techniques to generate the final devices. For the single cell experiments, devices with two different feature heights were required to mix the cAMP to provide temporally complex input stimuli (Figure 1-figure supplement 1a) (Stroock et al, 2002) and for these silicon masters we used two-step photolithography (Anderson et al, 2000). Briefly, to create the main cell channel an initial (160 µm thick) layer of SU-8 photoresist was spin coated, exposed to UV light, and developed on a silicon wafer. Subsequently, a second (15 µm thick) layer of photoresist was spun on top of this layer, the first layer features aligned to the mask for this second layer, exposed, and developed. For the population experiments, an 0.8 mm tall aluminum Y-channel master was machined with 2 mm wide and 6.5 mm long input channels and a 3 mm wide and 19 mm long cell area.

For both microfluidic devices, the microchannels were formed out of poly(dimethylsiloxane) (PDMS) via replica molding. Specifically, a 10:1 ratio of PDMS prepolymer to catalyst was poured on top of the master of interest, baked for 50 minutes at 65 °C, and then cut to size and removed from the master. Access holes for tubing were created using a 1.5 mm biopsy punch prior to plasma bonding the slabs to glass coverslips. Devices were then baked for at least 1 hour at 65 °C to aid in restoring a hydrophobic surface to the PDMS.

**Cell Perfusion and cAMP Stimulation**
For experiments, vegetative cells were harvested at 1.5-3x$10^6$ cells/mL, washed, and shaken at 1-2x$10^7$ cells/mL in developmental buffer (DB; 10 nM K/$Na_2$ phosphate buffer, 2 mM $MgSO_4$, 0.2 M $CaCl_2$, pH 6.5) for 4 to 5 hours before plating inside a microfluidic device. Single cell microfluidic devices were seeded with 2000-4000 cells and cells were permitted to adhere to the glass for 10 minutes before a constant flow rate of 4 µL/min was initiated using syringe pumps (Fusion Touch 200, Chemyx). Population devices were seeded with cells to the desired coverage and allowed to rest for 10 minutes before initiating flow of 10-100 µL/min and imaging 2 mm from the population edge. Macrofluidic dishes were seeded with 0.01 ML (1 ML = 6600 cells/$mm^2$) of cells and allowed to rest for 10 minutes before initiating flow of 1 mL/min. Cells were maintained at 22 °C throughout imaging. For experiments where single cell cytosolic cAMP traces were taken from a population, 15-20% Epac1camps- and mRFPmars-expressing cells were mixed in the population to facilitate single-cell tracking.

Single cells in microfluidic chips have a higher initial average response than those in microfluidic dishes (Figure 1-figure supplement 1B, C) as the transition from 0 to the stimulus concentration of cAMP in microfluidic chips is sharp, whereas in macrofluidic dishes chaotic mixing of the cAMP with buffer leads to a more gradual transition.

**Image Acquisition**
Cells were observed using an inverted epifluorescence microscope (TE300, Nikon) equipped with a Xenon lamp, automated excitation and emission filter wheels (Ludl), automated stage (Ludl), and oil immersion objectives (20X UPlanSApo NA 0.85, Olympus and 60X Plan Apo NA 1.40, Nikon). For FRET measurements, three fluorescent images were taken at each time point and are described using the following notation: $F_{ex_\lambda Range, em_\lambda Range}^{CellLine}$ where $F$ is the fluorescence intensity, the cell line is the line expressing the donor fluorophore only (D, here ECFP), the acceptor fluorophore only (A, here EYFP), or the epac1camps complex (EPAC). The excitation wavelength filter ranges are 436/20 nm for D and 500/20 nm for A, and the emission filter ranges are 470/24 nm for D and 535/30 nm for A (ET series filters, Chroma). A dichroic long-pass filter (T455LP, Chroma) further separated eCFP excitation from emission fluorescence during imaging. Images were captured using a back-illuminated Electron Multiplying CCD (EMCCD) camera (iXon+ 897, Andor) with a depth of 16 bits, 256x256 resolution, and 1000X gain. To minimize photodamage, exposure times were limited to 80 ms at 20X magnification and 20 ms at 60X magnification, an ND8 filter was in the emission path, and each field of view was imaged every 15 s. Acquisition was controlled using a custom Java plugin in Micro-Manager (Edelstein et al, 2010) that continually centered the cell of interest in single cell experiments and maintained focus at the plane of interest.

**Data Analysis**
Image analysis was performed using custom MATLAB (MathWorks) routines. Images were binarized and for single-cell data, single-cell masks were generated by thresholding each fluorescent cell against its local background. After masking, the $F^{EPAC}_{ex_\lambda D, em_\lambda D}$, $F^{EPAC}_{ex_\lambda D, em_\lambda A}$, and $F^{EPAC}_{ex_\lambda A, em_\lambda A}$ image intensities were averaged across each cell at each time point to reduce noise. To calculate changes in the FRET efficiency, we used the $Ef_{DA}/\gamma$-based FRET method (Salonikidis et al, 2008). Briefly, $E$ is the FRET efficiency, $f_{DA}$ is the fraction of Epac1camps complexes in their bound states, and $\gamma$ is the relative donor/acceptor extinction. This method improves on the traditional ratiometric FRET method, which uses $F^{EPAC}_{ex_\lambda D, em_\lambda A}/F^{EPAC}_{ex_\lambda D, em_\lambda D}$ as a readout of FRET efficiency, by correcting for any photobleaching that occurs during long experiments. Furthermore, this method is independent of pH changes unlike the ratiometric method (Salonikidis et al, 2008), meaning that any cytosolic pH changes that may occur do not affect our readout of the FRET efficiency. To find $Ef_{DA}/\gamma$, we use the formula

$$Ef_{DA}/\gamma = \frac{F^{EPAC}_{ex_\lambda D, em_\lambda A} - \alpha F^{EPAC}_{ex_\lambda A, em_\lambda A} - \beta F^{EPAC}_{ex_\lambda D, em_\lambda D}}{\alpha F^{EPAC}_{ex_\lambda A, em_\lambda A}} = f([cAMP]) \tag{S1}$$

where $\alpha = F^{A}_{ex_\lambda D, em_\lambda A}/F^{A}_{ex_\lambda A, em_\lambda A}$, the relative acceptor fluorescence signal, and $\beta = F^{D}_{ex_\lambda D, em_\lambda A}/F^{D}_{ex_\lambda D, em_\lambda D}$, the donor bleedthrough. On our imaging setup, $\alpha = 0.054$ and $\beta = 0.906$. As Epac1camps is in a low FRET configuration when bound to cAMP and a high FRET configuration when unbound, we use $-Ef_{DA}/\gamma$ as our FRET intensity here. All single cell and population average FRET intensities are normalized to the baseline levels observed for either the single cell or the population at the beginning of each experiment prior to stimulation of the cells or synchronization of the population. FRET signal units are defined as the change in $-Ef_{DA}/\gamma$ where $E$ is the FRET efficiency, $f_{DA}$ is the fraction of Epac1camps complexes in their bound states, and $\gamma$ is the relative donor/acceptor extinction and 0 is defined as the baseline value of $-Ef_{DA}/\gamma$ for the experiment when cells are in cAMP-free buffer. When analyzing population average firing rates, spikes were required to be 0.3 FRET signal units in height.

**Simulations**
Simulations were done using the Euler-Maruyama method (Kloeden & Platen, 1992) with a time step $\Delta t = 0.005$, and started with random initial conditions. For phase diagrams calculations, longer equilibration periods were used to eliminate the effect of the initial conditions. Spikes were defined as peaks in activator with values above zero and rates were calculated by counting these spikes divided by simulation time.

Parameters used for all simulations throughout the entire manuscript (unless stated otherwise):
- $\epsilon = 0.1$ (ratio between the activator and repressor time scales)
- $\gamma = 0.5$ (repressor degradation rate)
- $c_0 = 1.2$ (steady-state repressor value in the absence of external cAMP)
- $\sigma = 0.15$ (noise strength)
- $N = 100$ (number of "cells" in population simulations)
- $a = 0.058$ (cAMP response magnitude)
- $\alpha_0 = 800$ (basal cAMP leakage rate)
- $\alpha_{PDE} = 10^3$ (basal PDE leakage rate)
- $K_d = 10^{-5}$ (cAMP response threshold)
- $S = 10^6$ (firing cAMP release rate)

**ACKNOWLEDGEMENTS**

We thank R. Cooper, S. Sawai, and C. Parent for contributing reagents, and E.C. Cox, A. Lang, D. Yi, H.G. Garcia, and N. Wingreen for helpful discussions and comments. This work was supported by NIH Grants P50 GM071508 and R01 GM098407, by NSF-DMR 0819860, by an NIH NRSA fellowship (AES), and by Searle Scholar Award 10-SSP-274 (TG). PM and JN were partially funded by NIH K25 GM086909. DJS was funded by K25 GM098875-02.

**AUTHOR CONTRIBUTIONS**

A.E.S, D.J.S., J.N., T.M., P.M., and T.G. conceived and designed the experiments, A.E.S. and T.G. performed the experiments, and A.E.S, D.J.S., J.N., T.M., P.M., and T.G. analyzed the data, contributed analysis tools, and wrote the paper.

**CONFLICT OF INTEREST**

The authors declare that no conflict of interest exists.

Mixer for Microchannels. *Science* **295**: 647-651

Szabó B, Szöllösi GJ, Gönci B, Jurányi Z, Selmeczi D, Vicsek T (2006) Phase transition in the collective migration of tissue cells: Experiment and model. *Physical Review E* **74**: 061908

Takeda K, Shao D, Adler M, Charest PG, Loomis WF, Levine H, Groisman A, Rappel W-J, Firtel RA (2012) Incoherent Feedforward Control Governs Adaptation of Activated Ras in a Eukaryotic Chemotaxis Pathway. *Science Signaling* **5**: ra2

Tomchik K, Devreotes PN (1981) Adenosine 3',5'-monophosphate waves in Dictyostelium discoideum: a demonstration by isotope dilution--fluorography. *Science* **212**: 443-446

Ullner E, Buceta J, Díez-Noguera A, García-Ojalvo J (2009) Noise-Induced Coherence in Multicellular Circadian Clocks. *Biophysical Journal* **96**: 3573-3581

Vasiev B, Siegert F, Weller II CJ (1997) A Hydrodynamic model for *Dictyostelium discoideum* Mound Formation. *Journal of Theoretical Biology* **184**: 441-450

Vasiev BN, Hogeweg P, Panfilov AV (1994) Simulation of *Dictyostelium Discoideum* Aggregation via Reaction-Diffusion Model. *Physical Review Letters* **73**: 3173-3176

Wang CJ, Bergmann A, Lin B, Kim K, Levchenko A (2012) Diverse Sensitivity Thresholds in Dynamic Signaling Responses by Social Amoebae. *Science Signaling* **5**: ra17

Waters CM, Bassler BL (2005) Quorum Sensing: Cell-to-Cell Communication in Bacteria. *Annual Review of Cell and Developmental Biology* **21**: 319-346

Youk H, Lim WA (2014) Secreting and Sensing the Same Molecule Allows Cells to Achieve Versatile Social Behaviors. *Science* **343**

Young JW, Locke JCW, Elowitz MB (2013) Rate of environmental change determines stress response specificity. *Proceedings of the National Academy of Sciences* **110**: 4140-4145

Zhang HP, Be'er A, Florin E-L, Swinney HL (2010) Collective motion and density fluctuations in bacterial colonies. *Proceedings of the National Academy of Sciences* **107**: 13626-13630

**FIGURE LEGENDS**

**Figure 1. Modeling cytosolic cAMP responses to external cAMP stimuli in individual *Dictyostelium* cells.** (A, B) Experimental observation of a bifurcation: Cytosolic cAMP responses to an externally applied cAMP stimulus of 1 nM (A) and 10 µM (B) at 5 minutes in three individual *Dictyostelium* cells expressing an Epac1camps-FRET sensor (cells stimulated using a custom microfluidics device (see Figure E2); FRET signal is a normalized ratiometric fluorescence intensity measurement proportional to cytosolically produced cAMP (Salonikidis et al, 2008); see Materials and Methods. (C) Schematic of proposed model (see text for details) (D) Cytosolic cAMP responses in single cells to successive externally applied cAMP stimuli of 1 nM followed by 2 nM (dark blue) or 10 nM (light blue) step. (E, F) Phase portraits for a small (E) and a large (F) step stimulus (corresponding to (A) and (B), respectively), with repressor (R) nullcline (ie. $\frac{dR}{dt} = 0$) shown in red and activator (A) nullclines (i.e. $\frac{dA}{dt} = 0$ ) shown in green (see text for model details). For the activator two nullclines are shown corresponding to a pre-stimulus (light green) and a post-stimulus (dark green) regime. A fixed point for the dynamics occurs where the S-shaped activator nullcline intersects the repressor nullcline (red line). The response trajectory is shown in black. See Figure E3 for details about the FHN model fixed point behavior. (G, H) Activator variable as a function of time for (G) a small stimulus (corresponding to (A, E)) and for (H) a large stimulus (corresponding to (B, F), green dashed line indicates stimulus onset). For simulated data throughout the paper, time unit "T" is defined as the average minimum period in the FHN model for a single cell and "Amplitude" is defined as the mean height of spikes at 1 nM "internal cAMP". Simulated model time courses will be shown in shades of green. Experimental time courses will be shown in shades of blue with time units of minutes and amplitude in arbitrary FRET signal units.

**Figure 2. Phenomenological agreement between model and experiments.** (A) Experimental mean accommodation spikes for externally applied cAMP stimuli of 1 nM (light blue), 100 nM (medium blue), and 10 µM (dark blue) (see main text for discussion). (B, C) Experimental (B) and modeled (C) mean initial accommodation spike widths. (D, E) Experimental (D) and modeled (E) mean oscillation times, with experimental mean oscillations found by identifying the peak Fourier transform. Error bars in (A) and (B) represent SEM, error bars in (D) represent errors by bootstrapping. Colored data points in (B) correspond to data in (A), with additional mean accommodation spike widths taken at 10 nM and 1 µM (n=16, 14, 14, 20, and 11 cells).

**Figure 3. Cytosolic cAMP responses depend on the rate of externally applied cAMP.** (A, B) Externally applied cAMP stimuli (black) with a step and an exponential ramp to a final height of 1 nM cAMP (A) and with a step, a small intermediate step to 1% of the final height (experimental data only), and an exponential ramp to a final height of 300 nM cAMP (time of experimental ramp denoted with grey background). (B). Corresponding activator variable (green) and experimental data for three independent cells (blues) as a function of time are shown below. Model traces are shown for two different degrees of cellular excitability, $\epsilon = 0.1$ (light green) and 0.2 (dark green), showing a diversity of responses similar to that seen experimentally. (C, D) Phase portraits for a small (C) and a large (D) exponential ramp stimulus (corresponding to (A) and (B), respectively); R nullcline shown in red, pre-stimulus A nullcline shown in light green, and post-stimulus A nullcline in dark green. The response trajectory is shown in black.

**Figure 4. Cytosolic cAMP responses are entrainable to external cAMP stimuli and have a refractory period.** (A) Cytosolic cAMP responses in single cells to externally applied 10 nM cAMP pulses of 1 minute (top) and 5 minutes (bottom) with a 6 minute period (green dashed lines indicate stimulus onsets, red dashed lines indicates stimulus conclusions). (B, C) Phase diagrams summarizing experimental (B) and simulated (C) responses to various pulse widths and periods. "Entrainment Quality" is the mean correlation coefficient between the first period response and subsequent responses and is represented in color. Red regions display high correlation, while blue regions have low correlation.

**Figure 5. Multicellular model reproduces population behaviors in varying extracellular environments.** A phase diagram showing the coordinated population firing rate spanning a range of cell densities and flow rates for the model (A) and experiments from Gregor, et al. (2010) (B), with the mean firing rate represented in color and white vertical lines indicate nonlinear breaks in the x axis. Firing rates can also be considered as a function of the ratio between cell density and flow rate, $\rho/J$, as predicted by the model (C) and shown experimentally in Gregor, et al. (2010) (D). Low flow rates are not plotted in (D) because in this regime the effect of extracellular PDE is non-negligible. Model firing rates are normalized to an arbitrarily high frequency (~1/30) to scale maximum values to 1.

**Figure 6. Population model predicts slowing and decoupling of intracellular cAMP oscillations in a population with increased external cAMP.** (A) Firing rate phase diagrams for single cells in a population and the population as a whole as predicted by our model as a function of background and firing-induced cAMP. (B) Average "internal cAMP" responses of single simulated cells within a population (greens) and the population mean (black) responses at Points I-IV in (A) as simulated by the model. (C) Firing rate of experimental cellular populations with increasing external cAMP as a function of flow rate for cells plated at >0.5 ML (1 ML = 6600 cells/mm$^2$), imaged 2 mm from population edge. All spikes at least 0.3 FRET signal units in height. (D) Experimental population average cytosolic cAMP levels for experiments in a microfluidic device with 10 µL/min flow for no externally applied cAMP, 10 nM (low), 20 nM (intermediate), and 100 nM (high) steps of externally applied cAMP. Depending on fluid flow rates and externally applied cAMP levels, populations oscillate, have slow synchronous oscillations, do not oscillate but randomly fire, or oscillate asynchronously (stimulus onset for all assays is at 60 min). See also Figure E4.

**Figure 7. Intracellular noise in the cAMP circuit drives observed population behaviors.** (A) Firing rate phase diagrams for single cells in a population (top) and the population as a whole (bottom) as predicted by the model with minimal noise ($\sigma = 0.01$) as a function of background and firing-induced cAMP. See Figure E4 for other noise-source cases. (B) Example single cell (blues) and population (black) cytosolic cAMP traces taken from dual-expressing Epac1camps/mRFPmars tracer cells for oscillating populations at ~0.4 ML density, 10 µL/min flow subjected to steps of 10 nM and 500 nM cAMP. (C) Mean standard deviations of the single-cell cytosolic cAMP levels for cells in a population subjected to a step stimulus of cAMP as shown in (B) from 10-60 min post-stimulus. Values are normalized to the mean standard deviation of cells exposed to a 10 nM external cAMP step to show the relative increase in stochastic variability; errors by bootstrapping. (D) Single cell cytosolic cAMP responses to eight

1 nM pulses, 6 minute period with 1 min long pulses (grey) and two 1 minute, two 30 second, two 20 second, and two 10 second pulses (blues).

**Figure E1. Cytosolic cAMP responses to external cAMP stimulus.** Cytosolic cAMP responses in three individual cells to either no externally applied cAMP (A) or an externally applied cAMP stimulus of 1 nM (B), 10 nM (C), 100 nM (D), 1 µM (E), or 10 µM (F) at 5 minutes.

**Figure E2. Microfluidic device characterization.** (A) Single cell microfluidic device with on-chip mixing. Buffer and cAMP are delivered through the inlets, mixed, and delivered to cells in the imaging area. Dark blue channels are 160 µm high and light blue mixing areas are 175 µm high. (B, C) Single cell input-output ratios of time- and cell-averaged FRET responses in (B) microfluidic chips and (C) macrofluidic dishes for stimuli ranging from 100 pM to 10 µM, quantified for the first 3, 5, and 30 minutes after stimulation. Output/Time is normalized to the output of cells exposed to a 1 nM cAMP step in a microfluidic chip over the first three minutes of response. Bars represent standard error of the mean.

**Figure E3. Fixed points and behavior in the FHN model.** Phase portrait with activator $A$ nullcline and repressor $R$ nullclines showing different stimulus S values produce different dynamics in the FHN model. At low values of S when the intersection of the nullclines is on the leftside of the $A$ nullcline's local minimum, the system has a single stable fixed point (left plot). At higher values of S when the intersection shifts to the right side of the local minimum (shown in dark green), this fixed point becomes unstable (right plot) and oscillations result (see Figure 1H for example).

**Figure E4. Intracellular stochasticity drives experimentally observed population-level behaviors.** Firing rate phase diagrams for single cells in a population and the population as a whole as predicted by our model as a function of background and firing-induced cAMP with (A) standard noise level, $\sigma = 0.15$, equilibration time = 1000, (B) low intracellular noise ($\sigma = 0.01$), equilibration time = 116000, (C) noise in $\frac{d[cAMP]_{ex}}{dt}$ case with $\sigma_A = 0.01$ for the noise in $\frac{dA}{dt}$ and $\sigma_S = 0.15$, where $\frac{d[cAMP]_{ex}}{dt} = \alpha_f + \rho\alpha_0 + \rho S \frac{1}{N}\sum_{i=1}^{N}\Theta(A_i) - D[cAMP]_{ex} + \eta_S$ with $\eta_S$ an additive Gaussian white noise with a mean of zero and STD $\sigma_S = 0.15$, and (D) heterogeneous population case where $\sigma = 0.01$, $K_1 = 10^{-4}$, $K_2 = 10^{-6}$, and for half the population $K_d = K_1$ and $K = K_2$ for the other half. Model firing rates are normalized to an arbitrarily high frequency (~1/30) to scale maximum values to 1.

Figure 1

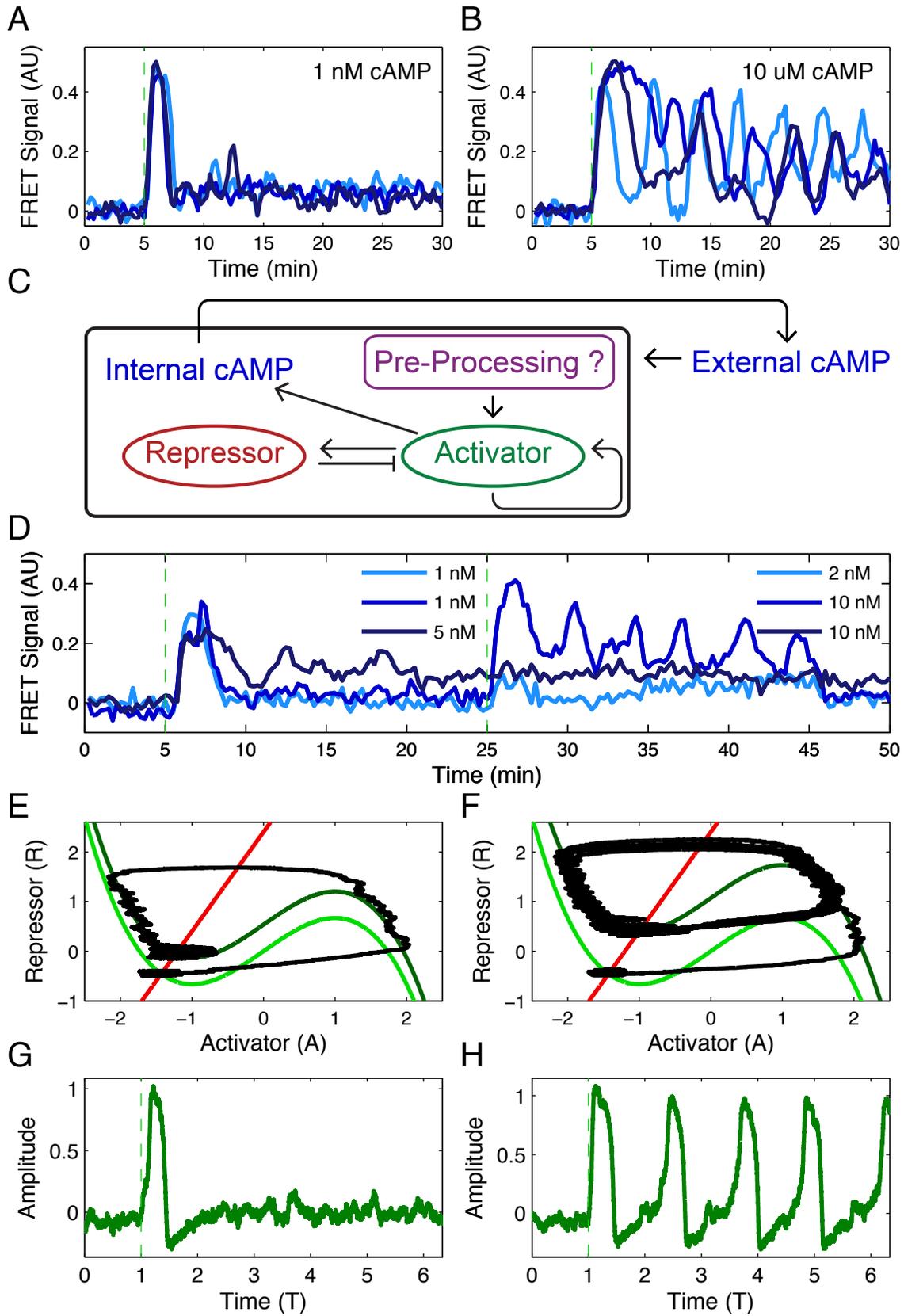

Figure 2

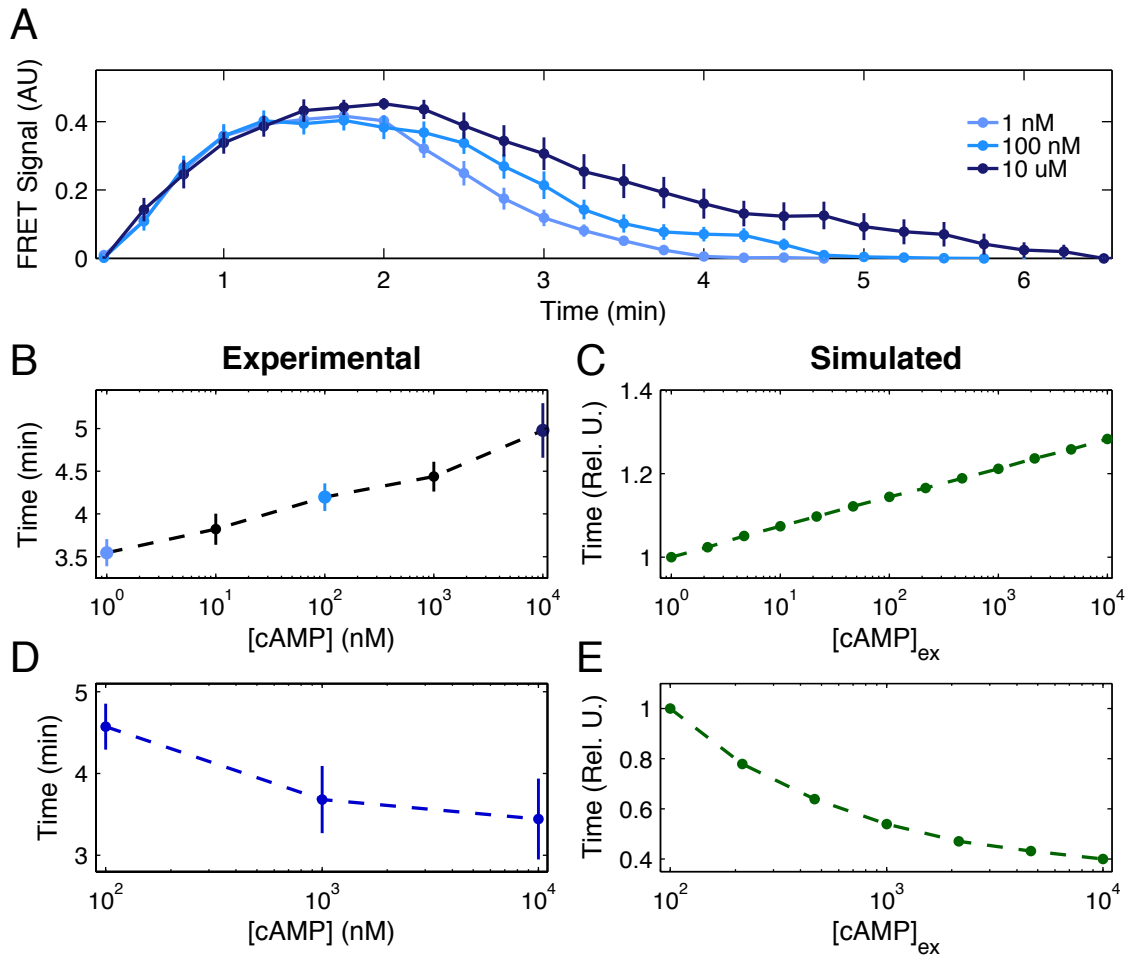

Figure 3

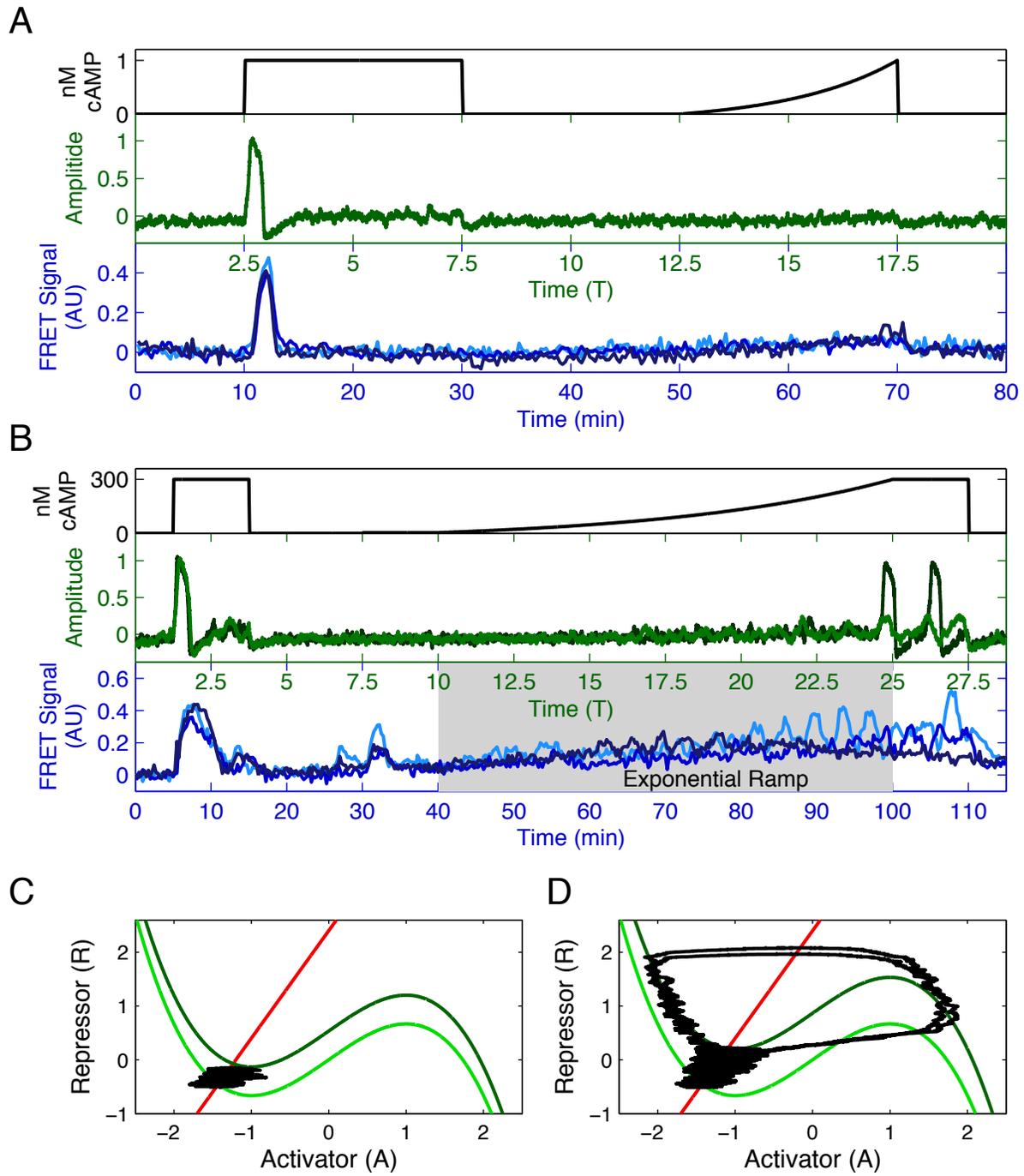

Figure 4

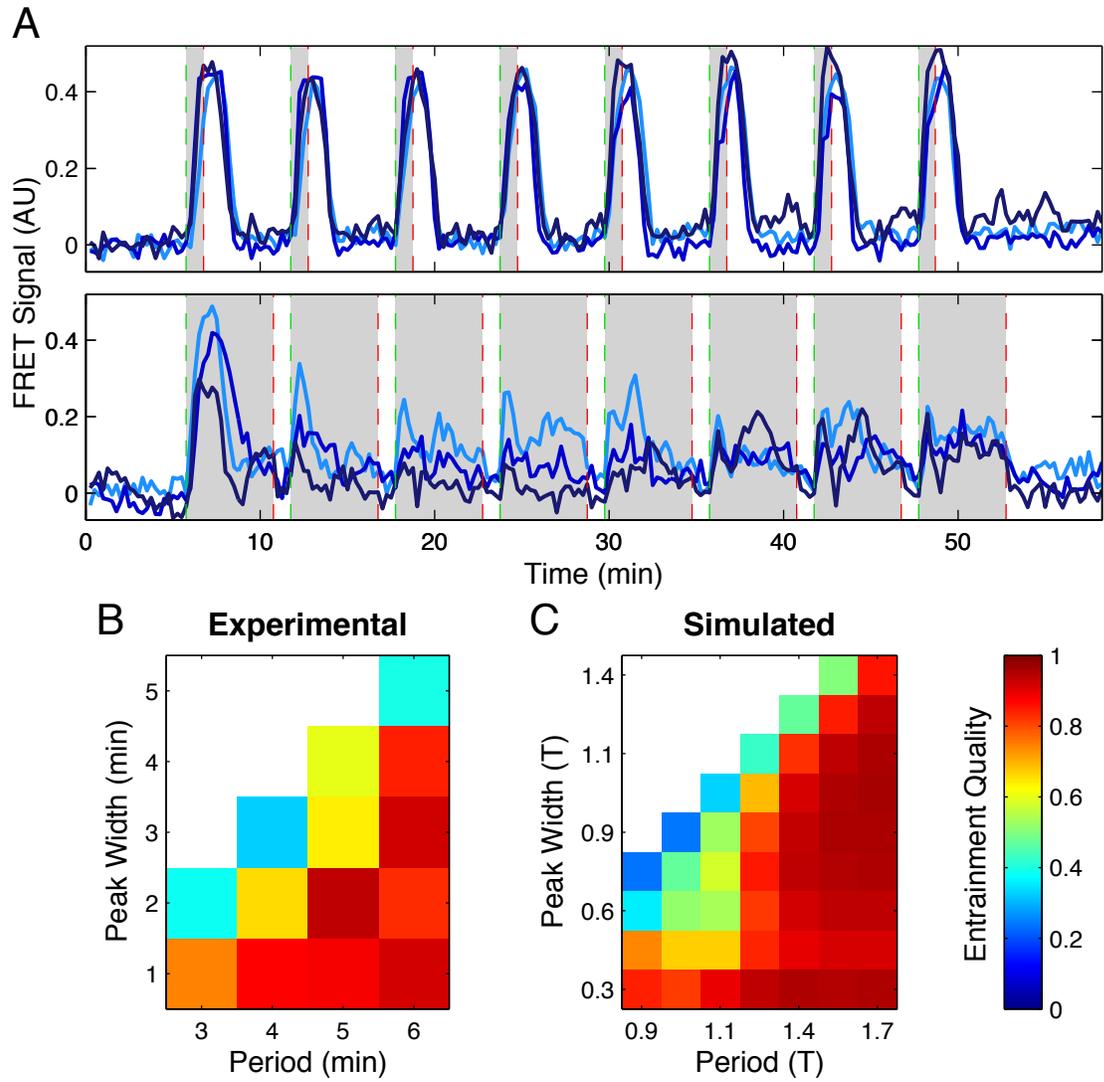

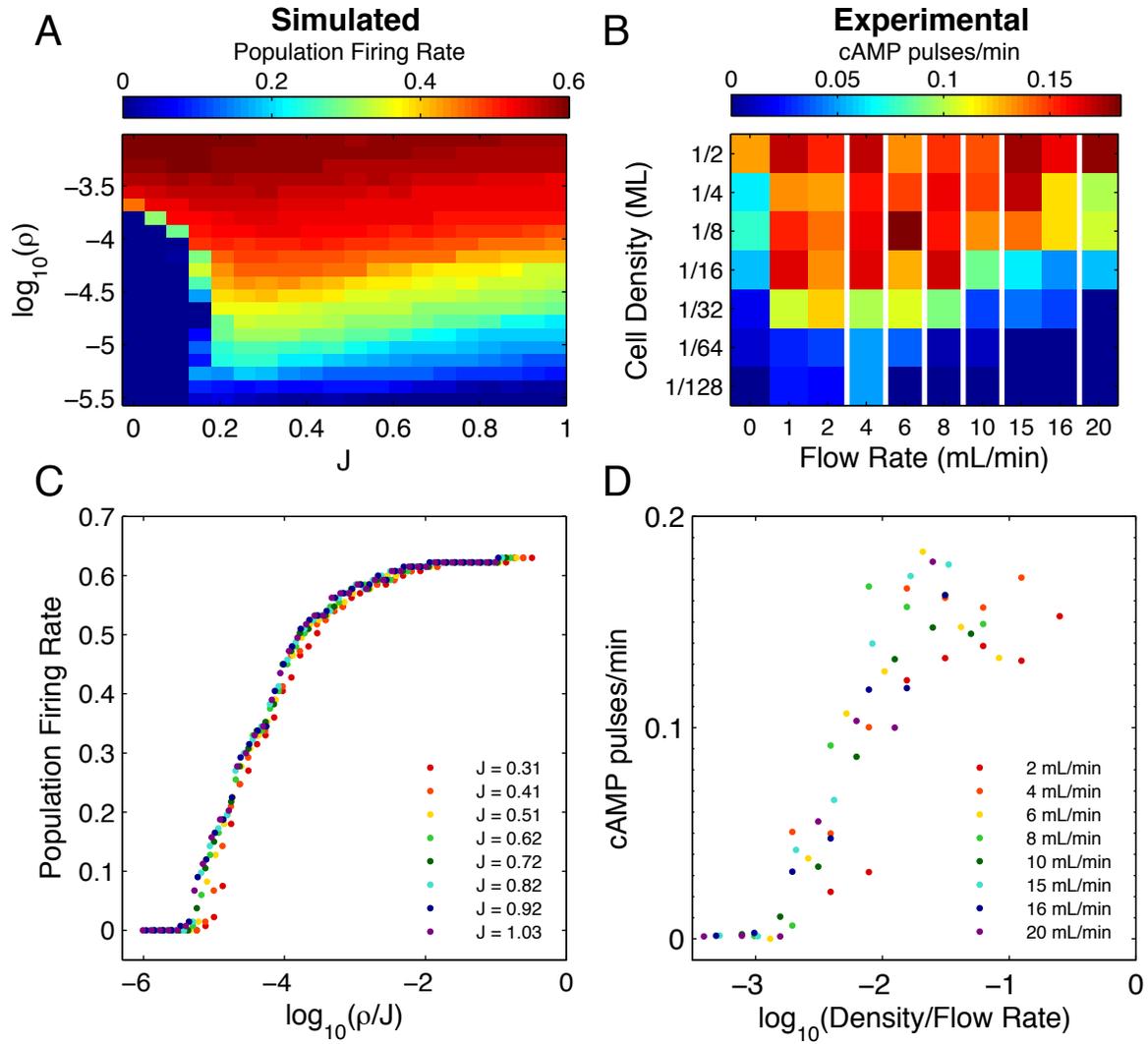

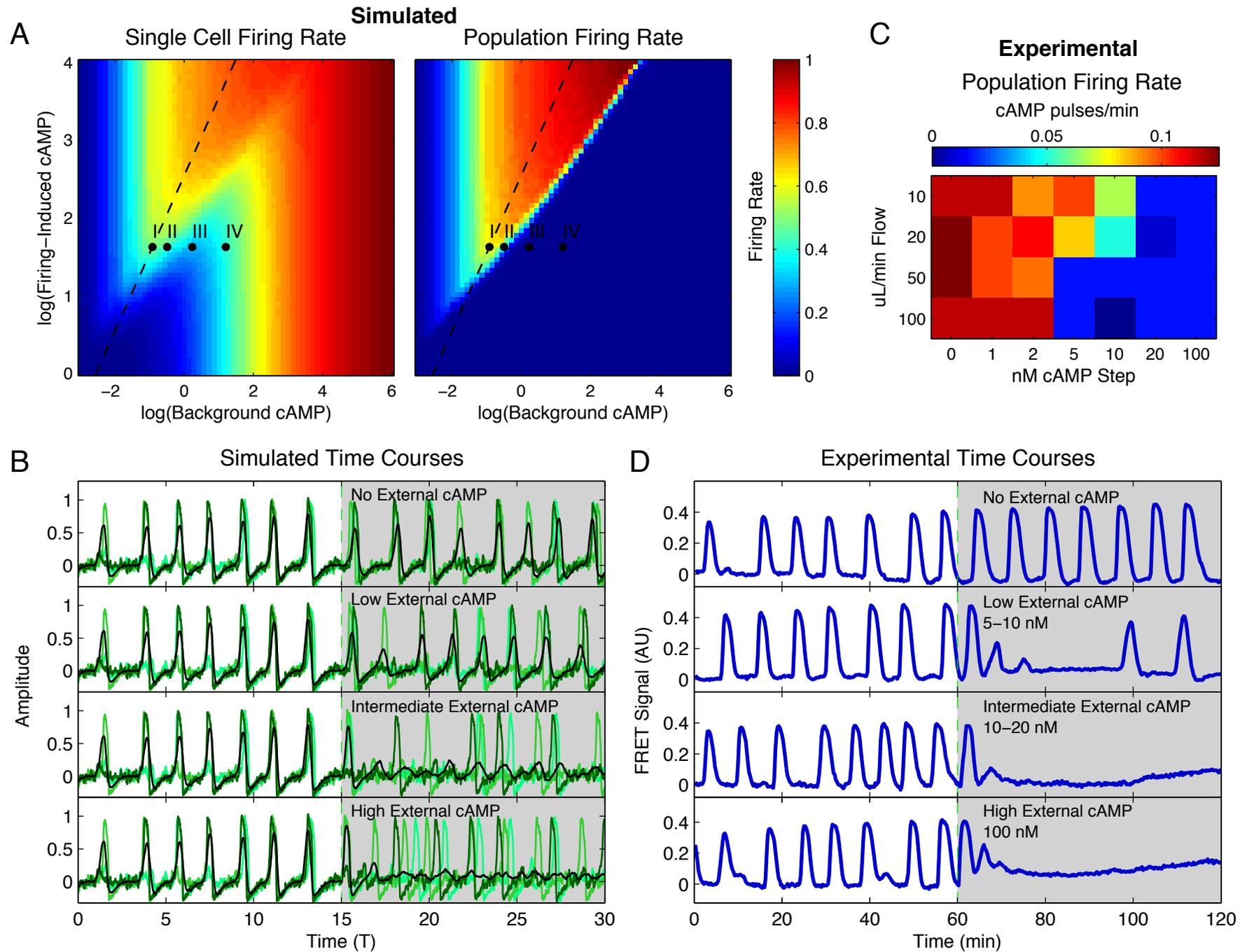

Figure 6

Figure 7

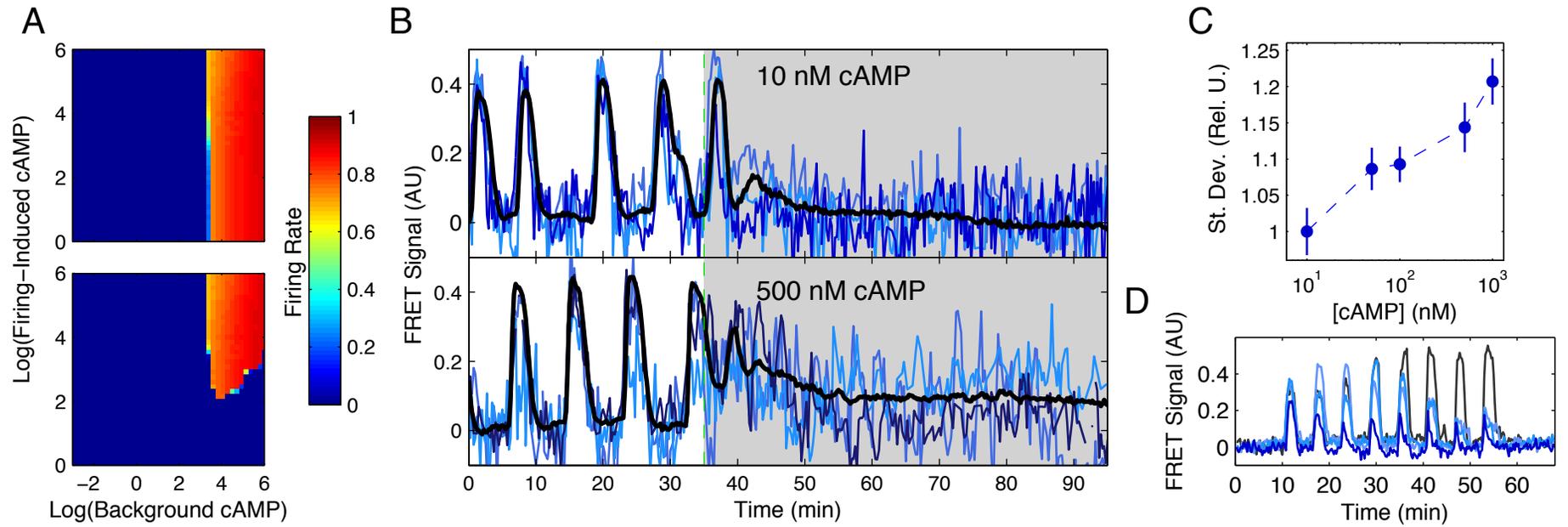

Figure E1

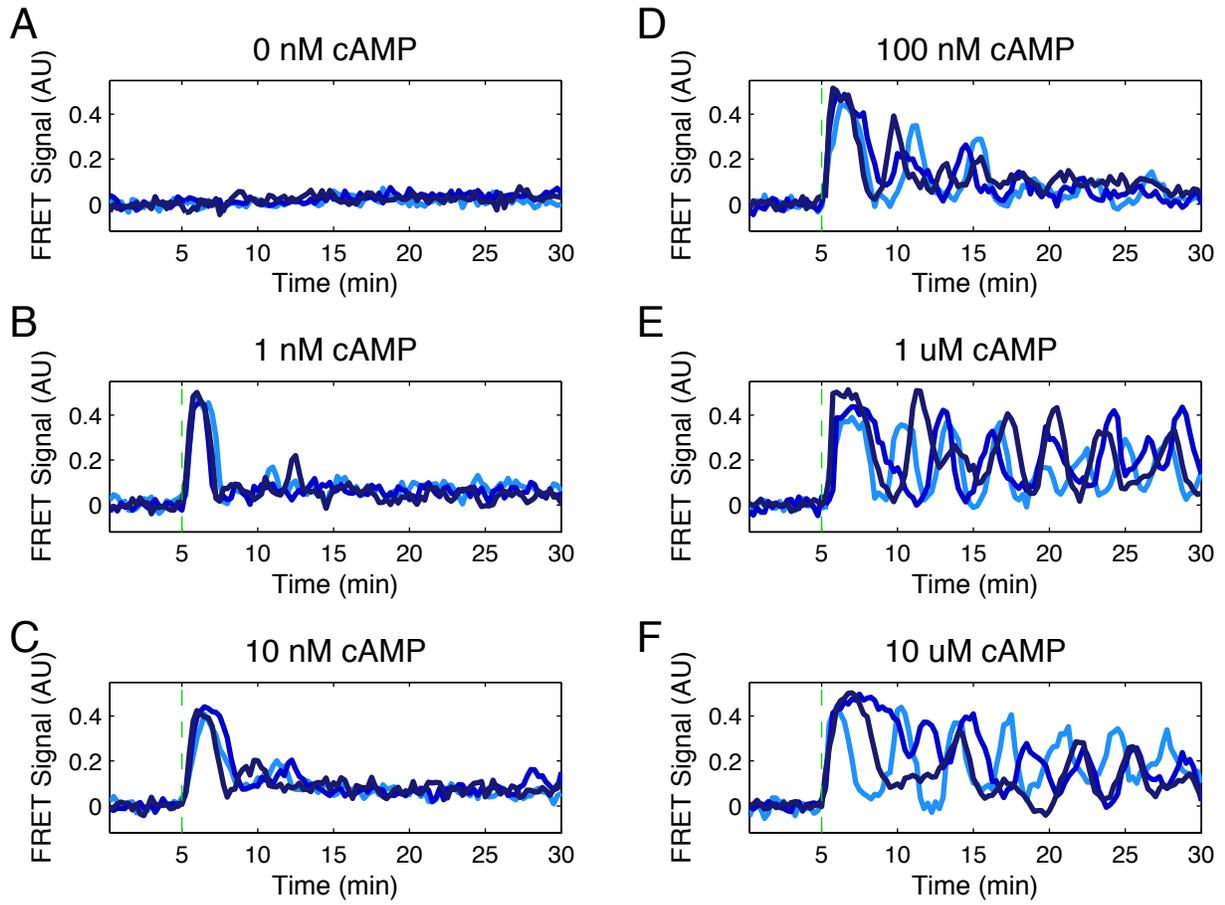

# Figure E2

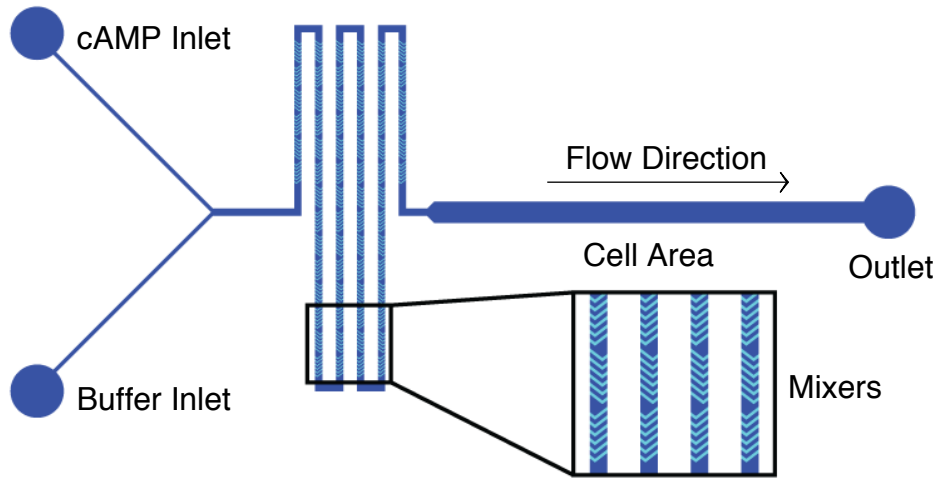

A

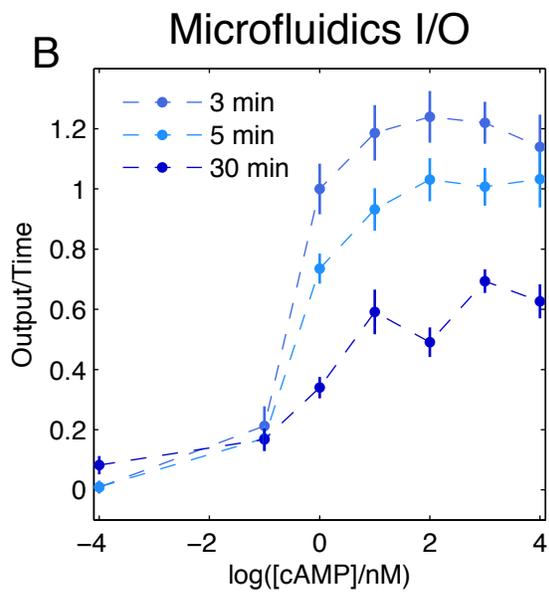

B

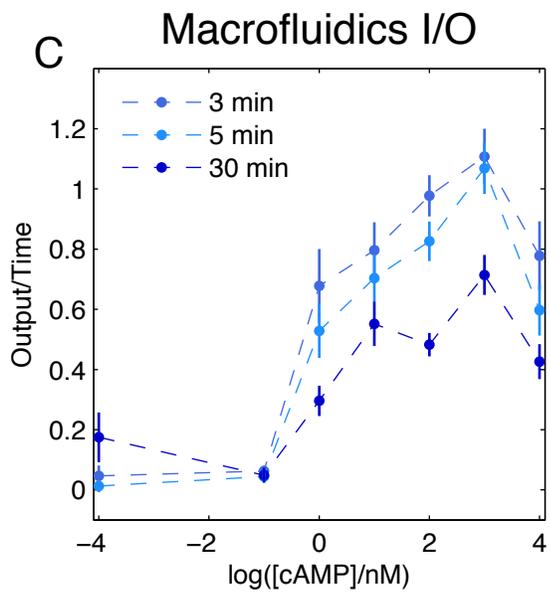

C

Figure E3

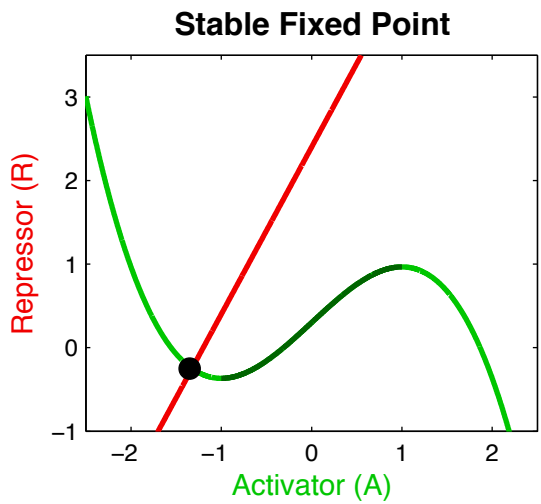
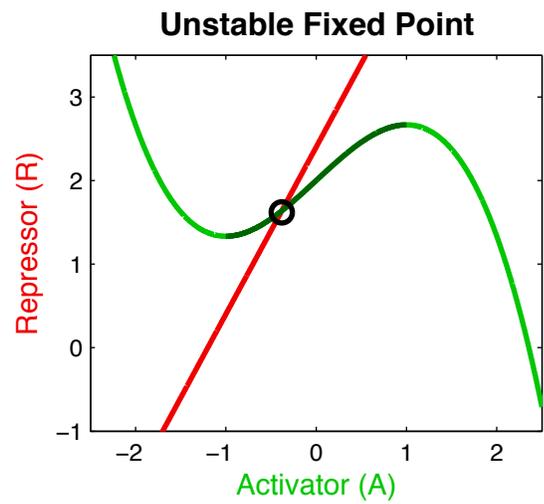

# Figure E4

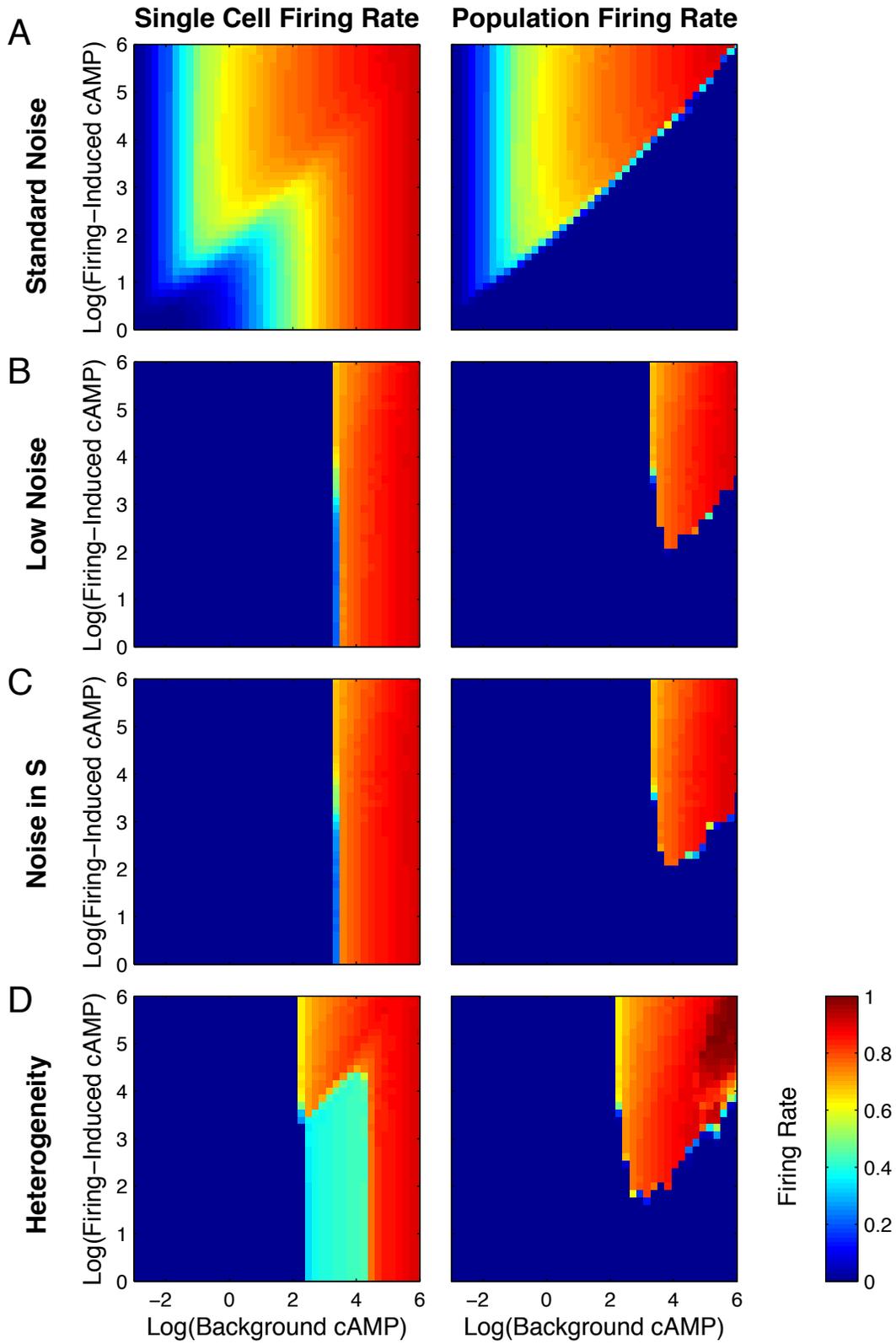